\documentclass[aip,jcp,amsmath,10pt,twocolumn,superscriptaddress,showpacs]{revtex4-1}
\usepackage[dvipdfmx]{graphicx}
\usepackage{color}
\usepackage{multirow}

\begin{document}

\title{Compositional descriptor-based recommender system accelerating the materials discovery}
\author{Atsuto \surname{Seko}}
\email{seko@cms.mtl.kyoto-u.ac.jp}
\affiliation{Department of Materials Science and Engineering, Kyoto University, Kyoto 606-8501, Japan}
\affiliation{Center for Elements Strategy Initiative for Structure Materials (ESISM), Kyoto University, Kyoto 606-8501, Japan}
\affiliation{JST, PRESTO, Kawaguchi 332-0012, Japan}
\affiliation{Center for Materials Research by Information Integration, National Institute for Materials Science, Tsukuba 305-0047, Japan}
\author{Hiroyuki \surname{Hayashi}}
\affiliation{Department of Materials Science and Engineering, Kyoto University, Kyoto 606-8501, Japan}
\affiliation{JST, PRESTO, Kawaguchi 332-0012, Japan}
\affiliation{Center for Materials Research by Information Integration, National Institute for Materials Science, Tsukuba 305-0047, Japan}
\author{Isao \surname{Tanaka}}
\affiliation{Department of Materials Science and Engineering, Kyoto University, Kyoto 606-8501, Japan}
\affiliation{Center for Elements Strategy Initiative for Structure Materials (ESISM), Kyoto University, Kyoto 606-8501, Japan}
\affiliation{Center for Materials Research by Information Integration, National Institute for Materials Science, Tsukuba 305-0047, Japan}
\affiliation{Nanostructures Research Laboratory, Japan Fine Ceramics Center, Nagoya 456-8587, Japan}

\date{\today}

\begin{abstract}
Structures and properties of many inorganic compounds have been collected historically.
However, it only covers a very small portion of possible inorganic crystals, which implies the presence of numerous currently unknown compounds.
A powerful machine-learning strategy is mandatory to discover new inorganic compounds from all chemical combinations.
Herein we propose a descriptor-based recommender-system approach to estimate the relevance of chemical compositions where stable crystals can be formed [i.e., chemically relevant compositions (CRCs)].
As well as data-driven compositional similarity used in the literature, the use of compositional descriptors as a prior knowledge can accelerate the discovery of new compounds.
We validate our recommender systems in two ways.
Firstly, one database is used to construct a model, while another is used for the validation.
Secondly, we estimate the phase stability for compounds at expected CRCs using density functional theory calculations.
\end{abstract}

\maketitle
\section{Introduction}
Since Von Laue and Bragg father-son conducted pioneering X-ray diffraction studies over a century ago, many atomic structures of inorganic crystals have been collected.
Of the few available databases for inorganic crystal structures \cite{bergerhoff1987crystal,xu2011inorganic,downs2003american}, the Inorganic Crystal Structure Database (ICSD) \cite{bergerhoff1987crystal} contains approximately $5\times10^4$ inorganic crystals, excluding duplicates and incompletes.
Although this is a rich heritage of human intellectual activities, it only covers a very small portion of possible inorganic crystals.
Considering 82 non-radioactive chemical elements, the number of simple chemical compositions up to ternary compounds A$_a$B$_b$C$_c$ with integers $\max(a, b, c) \leq 15$ is $6 \times 10^7$, and increases to $5 \times 10^9$ for quaternary compounds A$_a$B$_b$C$_c$D$_d$.
It is true that many such chemical compositions do not form stable crystals, but the huge difference between these numbers implies the presence of numerous currently unknown compounds.
Conventional experiments alone cannot fill this gap.
Often, first principles calculations are used as an alternative approach \cite{jain2013commentary,curtarolo2012aflowlib,saal2013materials}, but systematic first principles calculations without a prior knowledge of the crystal structures are very expensive.

On the other hand, new compounds have been discovered by inspecting the similarity between chemical elements and their compositions.
Similarity has been measured using heuristic quantities such as the proximity in the periodic table, electronegativity, ionicity, and ionic radius.
These quantities are derived from either simplified theoretical considerations or chemists' intuition.
Based on the similarity between given compositions and entries of experimental databases for existing crystals, currently unknown chemically relevant compositions (CRCs) may be discovered.
However, a manual or intuitive approach has serious limitations.
A manual search typically handles only a few heuristic quantities at most.
An expected CRC can be given individually using known CRCs in a database.
In addition, a figure of merit is not provided because the expected CRCs are given qualitatively.

\begin{figure*}[tbp]
\begin{center}
\includegraphics[width=\linewidth,clip]{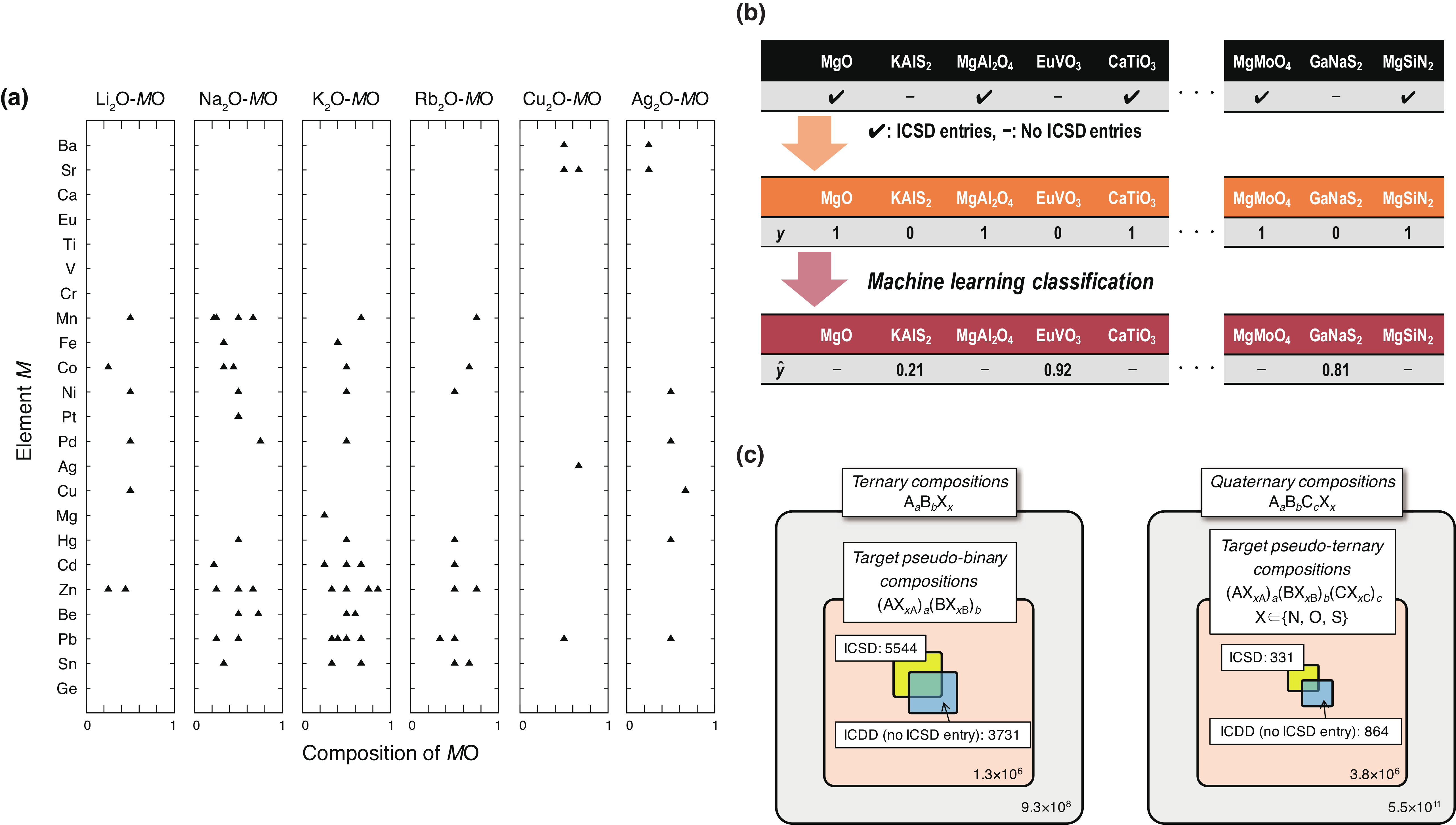} 
\caption{
(a) ICSD entries in $A_2$O-$M$O pseudo-binary systems, where $A$ and $M$ denote monovalent and divalent cations, respectively. 
Closed triangles indicate the compositions of the ICSD entries. 
Divalent cations are indicated in the order of Mendeleev number conceived by D.G. Pettifor \cite{pettifor1986structures}.
(b) Flowchart to find the expected CRCs using a classification technique. 
The expectant probability, $\hat y$, is estimated as a function of the set of descriptors by machine learning classifications.
The function gives the expectant probability for each ``no-entry'' composition.
A ``no entry'' composition with a high expectant probability is regarded as an expected CRC.
(c) Summary of the number of compositions included in our dataset.
}
\label{recommend1:Fig1}
\end{center}
\end{figure*}

As an example, consider searching for CRCs in Li$_2$O-$M$O ($M$: divalent cation) pseudo-binary systems from known CRCs.
Figure \ref{recommend1:Fig1} (a) shows the compositions in both Li$_2$O-$M$O and $A_2$O-$M$O ($A$; monovalent cation) systems where ICSD entries exist.
Known CRCs are widely scattered and depend on both elements $M$ and $A$, suggesting that cationic similarity may identify many currently unknown CRCs.
However, a quantitative figure of merit cannot be given without using machine learning (ML)-based methods.
We propose a novel ML-based approach to provide a quantitative figure of merit for numerous compositions.
The figure of merit for an expected CRC is given by ML-based modeling from known CRCs, which is hereafter called ``expectant probability.'' Many compositions with different degrees of expectant probabilities can be predicted as expected CRCs.

A few ML approaches to search for expected CRCs from inorganic databases have been proposed, where the expectant probabilities for candidate compositions were obtained using the compositional similarity on the basis of the entries in the database itself \cite{hautier2010finding,hautier2010data,seko2017matrix}.
Also, a ML model with respect to only the composition, which was estimated from a DFT database of the formation energies, predicts currently unknown CRCs\cite{PhysRevB.89.094104}.
A ML approach to predict compositions where glasses are formed was also proposed \cite{ward2016general}.

This study adopts a ML approach using compositional similarity defined by descriptors obtained from a set of well-known elemental representations. 
We demonstrate the potential possibility of descriptors for predicting currently unknown CRCs.
The present method can be regarded as a recommender system\cite{resnick1997recommender,aggarwal2016recommender}, which has become increasingly popular in a variety of scientific and non-scientific areas.
The present method corresponds to a kind of knowledge-based recommender system that utilizes prior knowledge about compositions.
A major advantage of knowledge-based recommender systems is avoiding the so-called ``cold-start'' problems.
In the present case, the cold-start problem is that the expectant probability cannot be estimated for a given composition due to the lack of related known CRCs.
As a result, few expected CRCs are recommended.
This may occur in applications to multicomponent systems such as a pseudo-ternary system where few known CRCs exist, which are the most interesting application of ML-based methods.
Therefore, the use of a knowledge-based method should contribute significantly to multicomponent expected CRCs.

\section{Methodology}
\subsection{Descriptors}
Herein, the compositional similarity is defined by a set of 165 descriptors composed of means, standard deviations, and covariances of established elemental representations \cite{PhysRevB.95.144110}, which is also similar to descriptors used in the literature\cite{ward2016general}.
This set of descriptors can cover a wide range of compositions.
Structural representations were not used because crystal structures with ``no entry'' compositions are unknown.
We adopted 22 elemental representations: (1) atomic number, (2) atomic mass, (3) period and (4) group in the periodic table, (5) first ionization energy, (6) second ionization energy, (7) electron affinity, (8) Pauling electronegativity, (9) Allen electronegativity, (10) van der Waals radius, (11) covalent radius, (12) atomic radius, (13) pseudopotential radius for the s orbital, (14) pseudopotential radius for the p orbital, (15) melting point, (16) boiling point, (17) density, (18) molar volume, (19) heat of fusion, (20) heat of vaporization, (21) thermal conductivity, and (22) specific heat.

\subsection{Knowledge-based recommender systems}
\begin{figure*}[tbp]
\begin{center}
\includegraphics[width=\linewidth,clip]{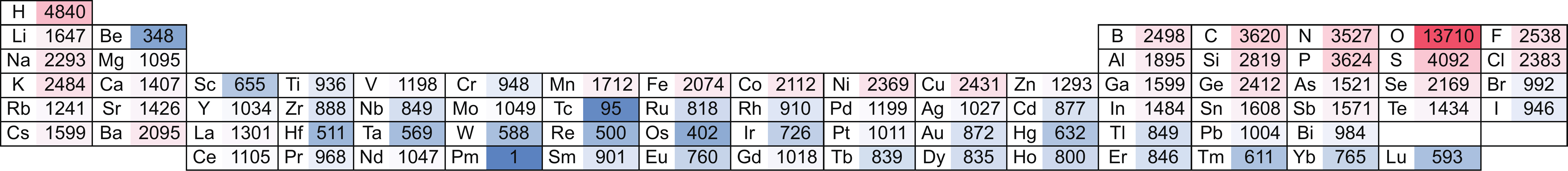} 
\caption{
Number of ICSD entries including each element.
Only entries where partial occupancy is not reported for any sites are considered.
This figure implies that inorganic-compound databases are biased to common metals, their intermetallics and oxides.
}
\label{recommend1:Fig2-icsd_analysis1}
\end{center}
\end{figure*}

Figure \ref{recommend1:Fig1} (b) schematically illustrates the practical aspects of the present method.
The expectant probability is estimated on the basis of a ML two-class classification, where responses have two distinct values of $y = 1$ and 0.
Since a supervised or semi-supervised classification approach requires a dataset with observations for both responses $y = 1$ and 0 (i.e., datasets for CRCs as well as non-existent compositions), we initially labeled compositions based on the criterion of whether the composition exists.
Entries in a crystal-structure database are regarded as $y = 1$ because they are known to exist.
On the other hand, it is not as simple to judge if a compound is non-existent at a given composition because the absence of a compound at a specific composition in the database (``no entries'') does not necessarily mean that the compound does not exist.
There are two reasons for a lack of entry.
1) A stable compound does not actually exist for a given composition.
2) The composition has not been well investigated.
It should be emphasized that inorganic-compound databases are biased to common metals, their intermetallics, and oxides (see Fig. \ref{recommend1:Fig2-icsd_analysis1}).
Few experiments have been devoted to other compositions.
In the present study, we simply assume that all ``no entry'' compositions are $y = 0$ in the training process.
Then the predicted response, $\hat y$, is regarded as the expectant probability.
Therefore, a ``no entry'' composition with high expectant probability is considered to be an expected CRC.

To date, many universal classifiers have been proposed in the field of ML, although they generate inconsistent boundaries for certain datasets.
We must choose an appropriate classifier with a high predictive power for the expected CRCs.
In this study, ``no entry'' compositions are used as both training and prediction data, which is contrary to the general classification problem where the prediction data is completely separate from the training dataset.
Thus, we have to avoid the case where a classifier predicts the response to be exactly zero for most ``no entry'' compositions.
To distinguish the ``no entry'' compositions from each other by the expectant probability, a classifier should be chosen to provide a continuous probability for ``no entry'' compositions.
We adopt three kinds of classifiers: logistic regression \cite{cox1958regression,hastieelements}, gradient boosting \cite{breiman1997arcing}, and random forest classifiers \cite{hastieelements,ho1995random}.

\begingroup
\squeezetable
\begin{table}[tbp]
\caption{
Elements and their charge-states included in ``no entries'' pseudo-binary and pseudo-ternary compositions used as candidates for new compositions where a stable compound is existent.
Elements included in candidate pseudo-ternary oxides and nitrides are shown as bold face.
}
\label{recommend1:TableElements}
\begin{ruledtabular}
\begin{tabular}{cl}
Charge state & Element \\
\hline
1$+$ & {\bf Li}, {\bf Na}, {\bf K}, Rb, Cs, Au, Cu, Hg, Tl, Ag \\
\hline
\multirow{2}{*}{2$+$} & Be, {\bf Mg}, {\bf Ca}, {\bf Sr}, {\bf Ba}, Ra, {\bf Zn}, Cd, Cu, Hg, Eu, Fe, \\
     & Co, {\bf Ni}, Pt, Ge, Sn, Pb, Ag, Ti, {\bf Pd}, V, Cr, Mn \\
\hline
     & {\bf Sc}, {\bf Y}, {\bf La}, Nd, Gd, Dy, Lu, B, {\bf Al}, {\bf Ga}, {\bf In}, Tl, Eu, \\
3$+$ & Fe, Co, Ni, Ce, Ru, Ir, P, As, {\bf Sb}, {\bf Bi}, Ag, Ti, Pd, \\ 
     & Nb, Ta, V, Mo, Cr, Mn \\
\hline
\multirow{2}{*}{4$+$} & {\bf Zr}, Hf, Tc, Os, Rh, Si, Pt, Ge, Sn, Pb, Ce, Ru, Ir,\\
     & {\bf Ti}, Pd, Nb, Ta, W, V, Mo, Cr, Mn, C \\
\hline
5$+$ & P, As, Sb, Bi, {\bf Nb}, Ta, W, V, Mo, Cr, N \\
\hline
6$+$ & Re, W, Mo, Cr, Mn, S, Se, Te \\
\hline
7$+$ & Re, Mn \\
\hline 
1$-$ & F, Cl, Br, I \\
\hline
2$-$ & O, S, Se, Te \\
\hline
3$-$ & N \\
\hline
4$-$ & C \\
\end{tabular}
\end{ruledtabular}
\end{table}
\endgroup

\subsection{Datasets}
The training dataset is composed of ``entries'' and ``no-entries'' in the ICSD.
The ``entries'' correspond to compounds up to septenary compositions.
Compounds reported to show a partial occupancy behavior are excluded.
Thus, the number of ``entries'' is 33,367.
The ``no-entries'' are used as the training data and the prediction data to find expected CRCs.
Figure \ref{recommend1:Fig1} (c) summarizes the number of compositions included in our dataset.
Although the present method is applicable to any kind of compound, we restrict the results to ionic compounds with normal cation/anion charge states.
Candidates of ``no entry'' pseudo-binary compositions A$_a$B$_b$X$_x$ are generated by considering combinations of $\{{\rm A, B, X}, a, b, x$\}.
We consider 930,142,080 chemical compositions expressed by integers satisfying the condition of $\max(a, b, x) \leq 15$.
Table \ref{recommend1:TableElements} shows candidates with elements A, B, and C and their charge-states.
Here, all charge-states are adopted whenever Shannon's ionic radii are reported.
Compositions that do not satisfy the charge neutrality condition of $n_{\rm A} a + n_{\rm B} b + n_{\rm X} x = 0$, are removed where $n_{\rm A}$, $n_{\rm B}$, and $n_{\rm X}$ denote the valences for elements A, B, and X, respectively.
Finally, ``entry'' compositions are removed from the set of candidate compositions.
Thereafter, 1,294,591 pseudo-binary compositions remain, which are used as the ``no entry'' data.
Additionally, pseudo-ternary oxides (AO$_{x_{\rm A}}$)$_a$(BO$_{x_{\rm B}}$)$_b$(CO$_{x_{\rm C}}$)$_c$, nitrides (AN$_{x_{\rm A}}$)$_a$(BN$_{x_{\rm B}}$)$_b$(CN$_{x_{\rm C}}$)$_c$, and sulfides (AS$_{x_{\rm A}}$)$_a$(BS$_{x_{\rm B}}$)$_b$(CS$_{x_{\rm C}}$)$_c$ are also considered to be ``no entry'' data.
Only a smaller number of elements and their charge states are adopted for pseudo-ternary compounds.
In all, there are 3,846,928 pseudo-ternary compositions.

\section{Results and discussion}
\subsection{Accuracy rates of classification models}

\begin{table}[htbp]
\caption{
Accuracy rates of three classification models. 
The number of ICDD entries found in samples of ``no entries'' pseudo-binary compositions according to the three classification models and the number of ICDD entries found by random sampling are also shown.
$N_{\rm ICDD} (s)$ denotes the number of ICDD entries in $s$ samples.
}
\label{recommend1:TableAccuracyRate}
\begin{ruledtabular}
\begin{tabular}{ccccc}
& Logistic &  Gradient    & Random &  \multirow{2}{*}{Random} \\
& regression &  boosting  & forest &   \\
\hline
Accuracy rate &  0.402 &  0.989 &  0.996 &  - \\
$N_{\rm ICDD} (100)$ & 6  & 22 & 33 & 0.29 \\
$N_{\rm ICDD} (1000)$ & 56 &  175 & 180 & 2.9 \\
\end{tabular}
\end{ruledtabular}
\end{table}

Here we show how the ML classification aids in the discovery of currently unknown CRCs.
Table \ref{recommend1:TableAccuracyRate} shows the accuracy rate of the binary responses by three classification models.
The accuracy rate indicates how well a classification model explains known CRCs.
A general definition of the accuracy rate was obtained from the number of accurate predictions both for $y = 1$ and $y = 0$ data. 
However, we assumed ``no entry'' compositions are $y = 0$ in the training process. 
Therefore, we defined the accuracy rate (AR) only for ICSD entries as 
\begin{equation}
{\rm (AR)} = \sum_{i=1}^{N_{\rm ICSD}} Y_i / N_{\rm ICSD},
\end{equation}
where $N_{\rm ICSD}$ and $Y_i$ denote the number of ICSD ``entry'' compositions and the predicted binary response for ICSD ``entry'' composition $i$, respectively.

Both the random forest and gradient boosting models have accuracy rates of almost unity, indicating that they can explain most known CRCs.
On the other hand, the accuracy rate of the logistic regression model is only 0.402, demonstrating that it cannot explain many known CRCs in terms of a binary response.
However, we emphasize that the classification model provides a non-zero expectant probability even for compositions with a zero response in terms of the binary response.

\subsection{Discovery rate of CRCs}

The power of discovering currently unknown CRCs is the most important feature of the classification model.
Therefore, we measured the efficiency for finding compositions of entries included in another inorganic compound database, Powder Diffraction File (PDF) by the International Centre for Diffraction Data (ICDD) \cite{ICDD}, from pseudo-binary and pseudo-ternary oxide ``no entry'' compositions of the ICSD.
Only 3,731 (0.3\%) of the 1,294,591 pseudo-binary (see Fig. \ref{recommend1:Fig1} (c)) and 842 (0.04\%) of the 1,933,994 pseudo-ternary oxide ``no entry'' compositions in the ICSD are included in the ICDD, respectively.
Therefore, it is obvious that discovering ICDD entries by random sampling is not effective.

\begin{figure}[tbp]
\begin{center}
\includegraphics[width=\linewidth,clip]{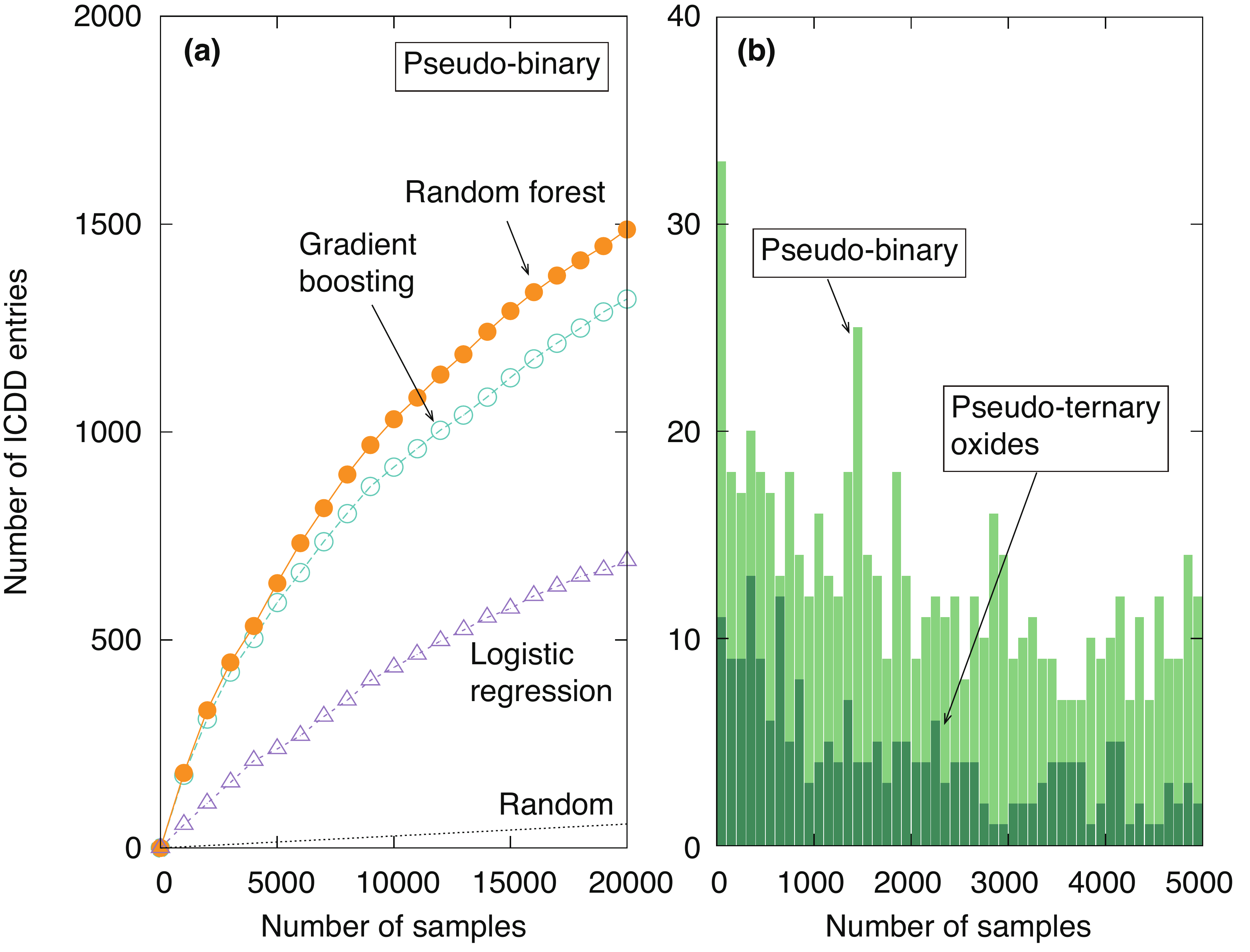} 
\caption{
(a) Number of ICDD entries found in ``no entries'' pseudo-binary compositions sampled by classification models and random sampling.
(b) Distributions of the increment of the number of ICDD entries as ``no entries'' pseudo-binary and pseudo-ternary oxide samples increase in the random forest model.
}
\label{recommend1:Fig3}
\end{center}
\end{figure}

Figure \ref{recommend1:Fig3} (a) shows the efficiency of the three classifiers for discovering ICDD entries from the pseudo-binary compositions.
Sampling of the ``no entry'' compositions is performed in decreasing order of the expectant probability.
Table \ref{recommend1:TableAccuracyRate} also shows the number of ICDD entries that can be discovered by sampling 100 and 1,000 ``no entry'' compositions.
Figure \ref{recommend1:Fig3} (b) shows the increment of the number of ICDD entries that can be discovered by the random forest model for pseudo-binary and pseudo-ternary oxide compositions.
When sampling 100 and 1,000 ``no entry'' pseudo-binary compositions according to the expectant probability ranking by the random forest model, which is the best among the three models, 33 and 180 compositions are found in the ICDD, respectively.
Even for pseudo-ternary oxide compositions, 11 and 85 compositions are found in the ICDD.
These discovery rates are approximately 250 and 190 times higher than that of random sampling (0.04\%), respectively, demonstrating that the ML classification approach significantly accelerates the discovery of currently unknown CRCs that are not present in the training database.
Figure \ref{recommend1:Fig3} (b) also indicates that the increment of the number of ICDD entries tends to decrease as the expectant probability decreases.
This provides evidence that the expectant probability can be regarded as a figure of merit for exploring currently unknown CRCs.

\begin{figure}[tbp]
\begin{center}
\includegraphics[width=\linewidth,clip]{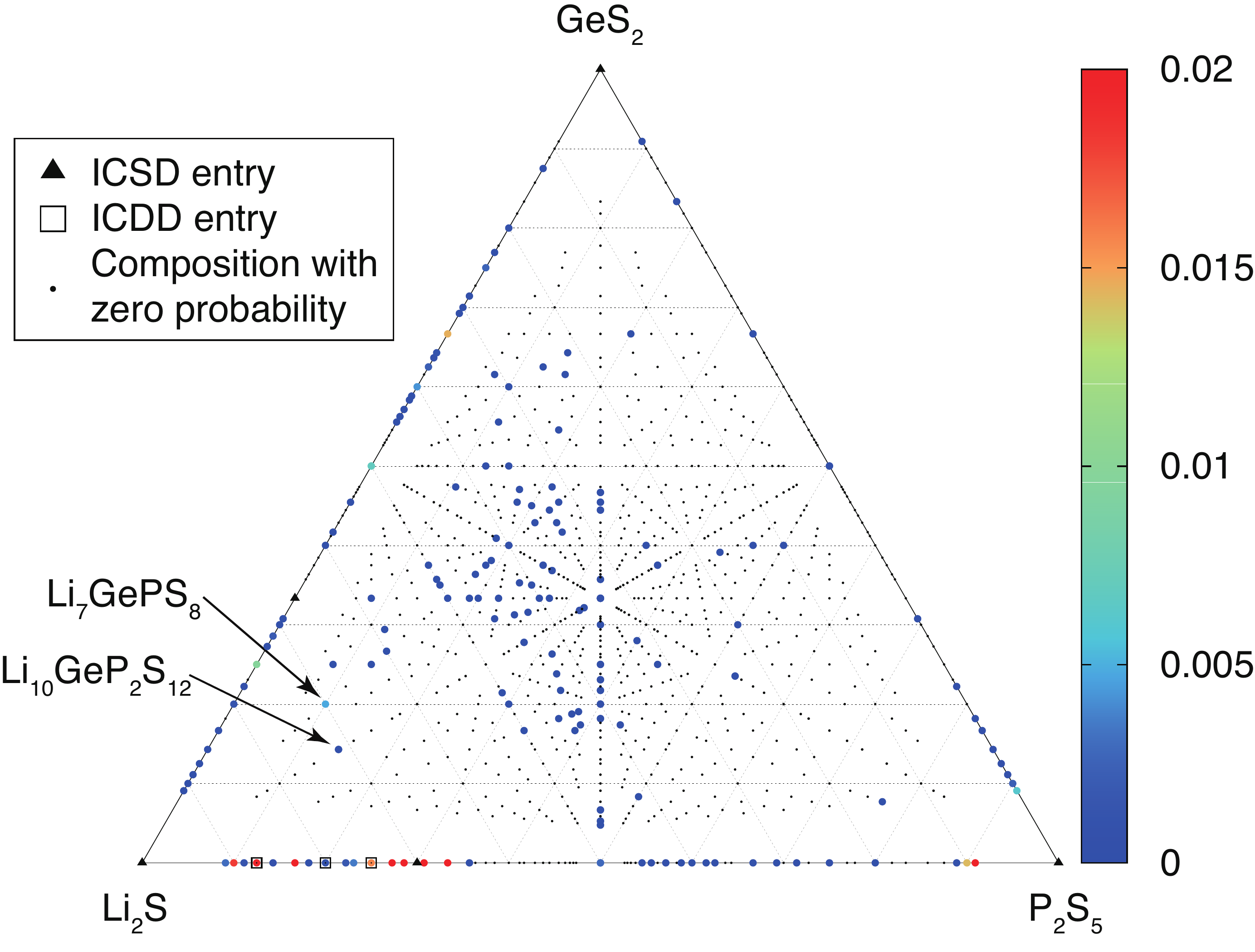} 
\caption{
Expectant probabilities of the compositions in Li$_2$S-GeS$_2$-P$_2$S$_5$ pseudo-ternary systems by the random forest model shown by the color palette along with the experimentally reported compositions.
Small closed circles show compositions with a zero expectant probability.
Closed triangles and open squares show ICSD entries without a partial occupancy and ICDD entries, respectively.
}
\label{recommend1:Fig4-Thio-LISICON}
\end{center}
\end{figure}

\subsection{Thio-LISICON pseudo-ternary system}

Another example demonstrating the predictive power of the ML prediction model can be found in the family of the Li$_2$S-GeS$_2$-P$_2$S$_5$ pseudo-ternary system, called thio-LISICON.
The Li$_2$S-GeS$_2$-P$_2$S$_5$ system shows high lithium-ion conductivity, and ICSD does not include any compounds only with full occupancy sites, which is often the case in pseudo-ternary systems.
Figure \ref{recommend1:Fig4-Thio-LISICON} shows the expectant probabilities of the compositions in the Li$_2$S-GeS$_2$-P$_2$S$_5$ system.
This system only has two experimentally reported compositions in the literature (i.e., Li$_{10}$GeP$_2$S$_{12}$ \cite{kamaya2011lithium} and Li$_7$GePS$_8$ \cite{kuhn2013tetragonal}).
Both are predicted to be expected CRCs.
In particular, Li$_7$GePS$_8$ shows the highest expectant probability in this pseudo-ternary system excluding pseudo-binary end members.
It should be emphasized that CRCs can be successfully predicted using the ML prediction model, even when there are no entry compositions in the corresponding pseudo-ternary system.

\subsection{Discovery of new compounds by DFT calculation}
\begin{figure}[tbp]
\begin{center}
\includegraphics[width=\linewidth,clip]{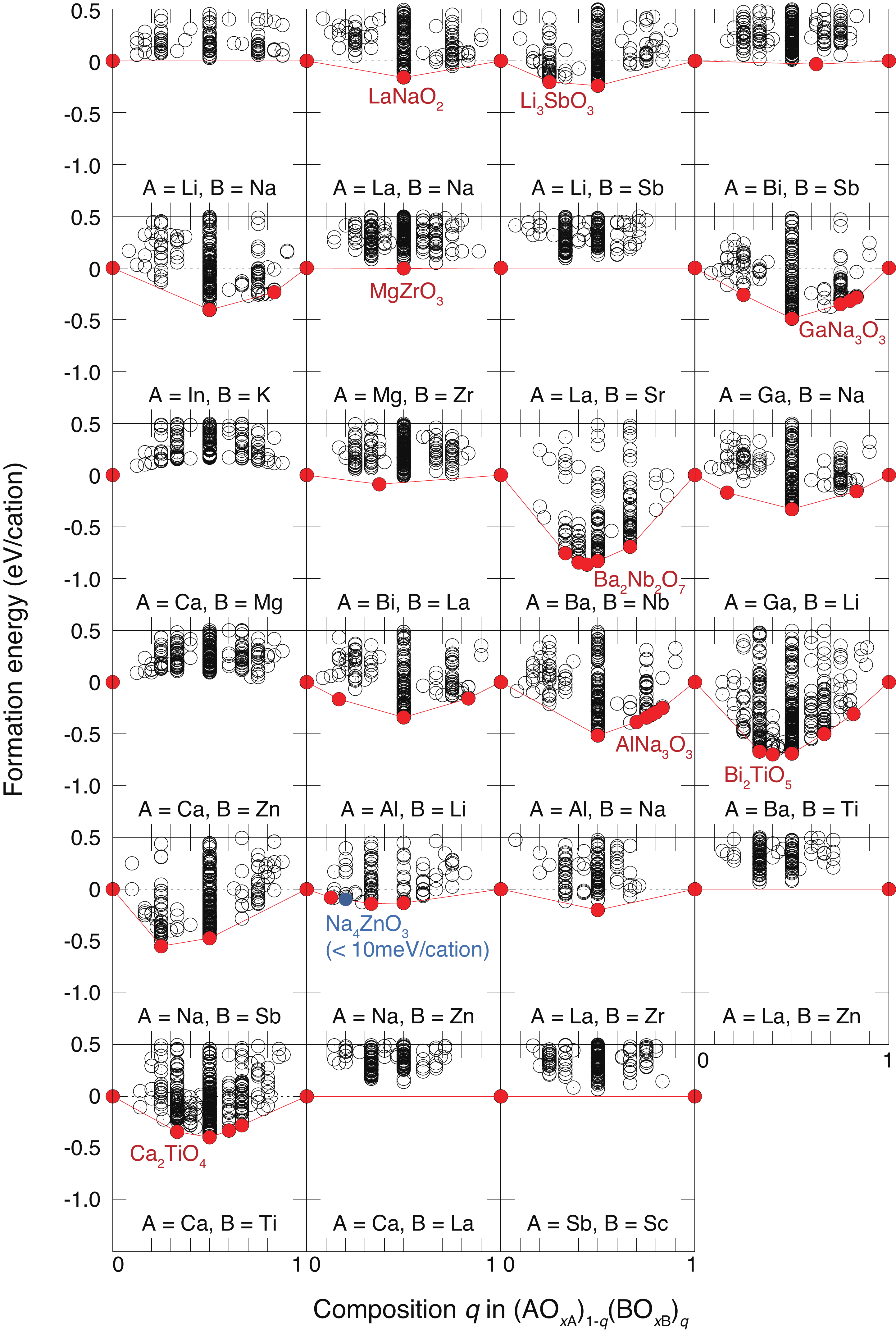} 
\caption{
DFT formation energies of prototype structures containing the expected CRCs shown in Table \ref{recommend1:TableDFT}.
Stable or nearly stable compounds corresponding to expected CRCs are highlighted.
}
\label{recommend1:Fig5-DFT}
\end{center}
\end{figure}

\begin{table}[htbp]
\caption{
Pseudo-binary expected CRCs by the random forest model.
Only expected CRCs where more than ten possible prototype structures are included in the ICSD are shown.
The lowest DFT energy among the prototype structures measured from the convex hull is also shown.
Compounds with an energy measured from the convex hull of less than 10 meV/cation are regarded as stable or nearly stable.
}
\label{recommend1:TableDFT}
\begin{ruledtabular}
\begin{tabular}{ccc}
\multirow{2}{*}{Composition} & Expectant & DFT energy from  \\ 
            & probability, $\hat y$ & convex hull (meV/cation) \\ 
\hline
LiNaO    &  0.234  & 29  \\
LaNaO$_2$   &  0.188  & \bf{0}   \\
Li$_3$SbO$_3$  &  0.148  & \bf{0}   \\
BiSbO$_3$   &  0.148  & 36  \\
InK$_3$O$_3$   &  0.146  & 22   \\
MgZrO$_3$   &  0.1424 & \bf{0}    \\
La$_2$SrO$_4$  &  0.124  & 98   \\
GaNa$_3$O$_3$  &  0.1    & \bf{0}    \\
CaMgO$_2$   &  0.098  & 169  \\
BiLaO$_3$   &  0.082  & 61   \\
Ba$_2$Nb$_2$O$_7$ &  0.082  & \bf{0}    \\
GaLi$_3$O$_3$  &  0.08   & 54   \\
CaZnO$_2$   &  0.078  & 94   \\
AlLi$_3$O$_3$  &  0.078  & 73   \\
AlNa$_3$O$_3$  &  0.074  & \bf{0}    \\
InNa$_3$O$_3$  &  0.068  & 76 \\
BaMgO$_2$   &  0.068  & 65 \\
Bi$_2$TiO$_5$  &  0.062  & \bf{0}  \\
Ba$_2$Ga$_2$O$_5$ &  0.062  & \bf{0}  \\
Al$_4$Y$_2$O$_9$  &  0.062  & 32 \\
NaSb$_3$O$_5$  &  0.058  & 96 \\
Na$_4$ZnO$_3$  &  0.056  & \bf{6}  \\
MgSrO$_2$   &  0.056  & 164    \\
K$_4$MgO$_3$   &  0.056  & 42 \\
La$_2$ZrO$_5$  &  0.054  & 21 \\
La$_2$ZnO$_4$  &  0.052  & 62 \\
Ca$_2$TiO$_4$  &  0.052  & \bf{0}  \\
CaTi$_3$O$_7$  &  0.052  & 154    \\
CaLa$_2$O$_4$  &  0.052  & 272    \\
SbScO$_3$   &  0.05   & 70 \\
\end{tabular}
\end{ruledtabular}
\end{table}

To discover currently unknown CRCs, density functional theory (DFT) calculations were used to examine the phase stability of pseudo-binary oxide systems containing compositions with a high expectant probability.
DFT calculations were performed for all ionic substitutions of the prototype structures included in ICSD.
The total number of DFT calculations was 19,338.
DFT calculations were performed using the plane-wave basis projector augmented wave (PAW) method \cite{PAW1,PAW2} within the Perdew--Burke--Ernzerhof exchange-correlation functional \cite{GGA:PBE96} as implemented in the VASP code \cite{VASP1,VASP2}.
The cutoff energy was set to 550 eV.
The total energy converged to less than 10$^{-3}$ meV.
The atomic positions and lattice constants were optimized until the residual forces were less than 10$^{-2}$ eV/\AA.

Table \ref{recommend1:TableDFT} shows 30 expected CRCs, which are not found in either ICSD or ICDD.
More than ten prototype structures are included in the ICSD for them.
We systematically computed the formation energy for all prototype structures, as listed in the ICSD.
Table \ref{recommend1:TableDFT} and Figure \ref{recommend1:Fig5-DFT} show the formation energy for the expected CRCs and the convex hull of the formation energy along with expectant probabilities for the pseudo-binary oxide systems, respectively.
The existence of stable or nearly stable compounds is predicted for ten of the expected CRCs: LaNaO$_2$, Li$_3$SbO$_3$, MgZrO$_3$, GaNaO$_3$, Ba$_2$Nb$_2$O$_7$, AlNa$_3$O$_3$, Bi$_2$TiO$_5$, Ba$_2$Ga$_2$O$_5$, Na$_4$ZnO$_3$, and Ca$_2$TiO$_4$.
Additionally, most of the other 20 compositions show small values of the energy measured from the convex hull, indicating that the expectant probability can be regarded as a figure of merit for exploring currently unknown CRCs.

\subsection{Elemental similarity}

\begin{figure}[tbp]
\begin{center}
\includegraphics[width=0.8\linewidth,clip]{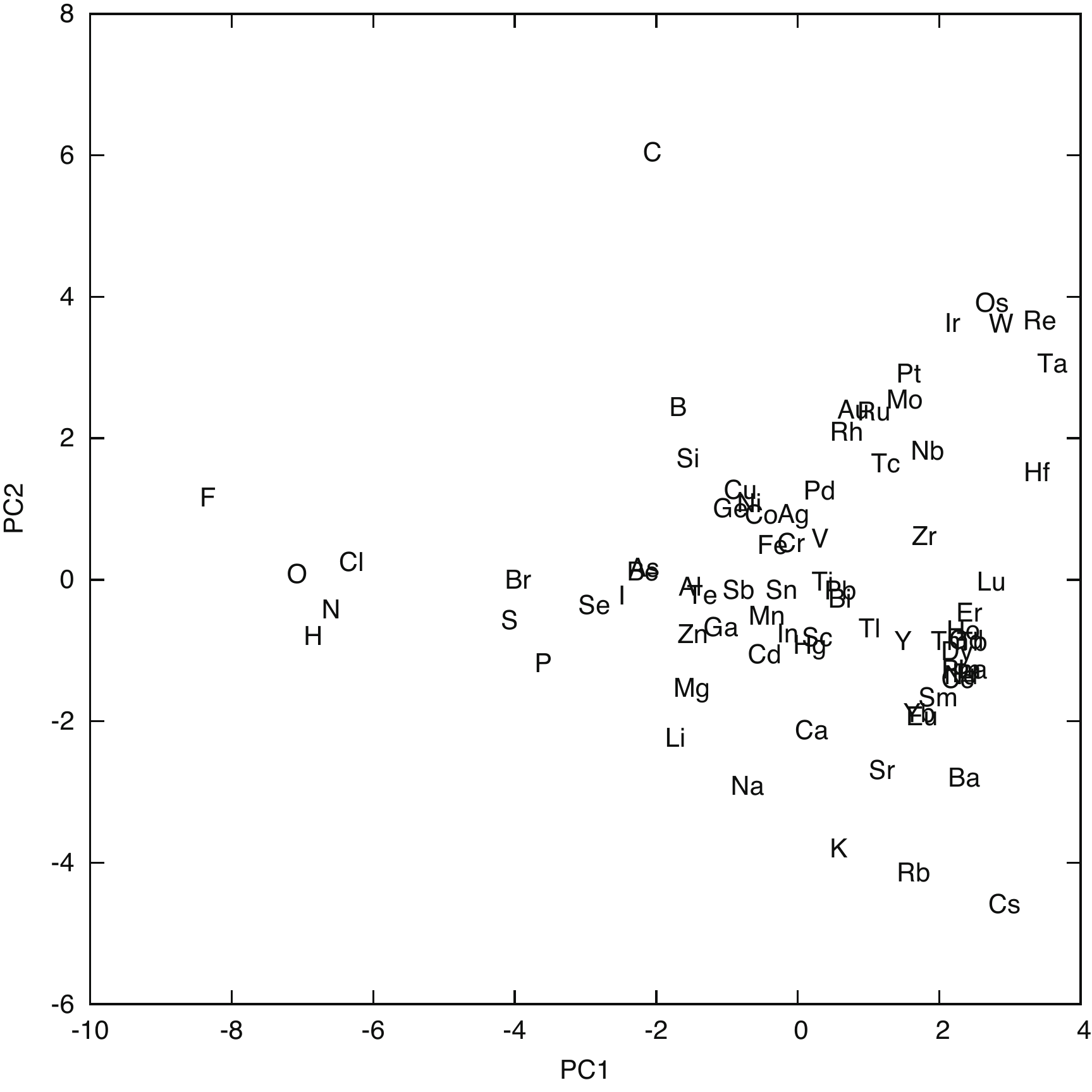} 
\caption{
Distribution of elements in the two-dimensional PC space. 
}
\label{recommend1:Fig6-PCAElement}
\end{center}
\end{figure}

\begin{figure*}[tbp]
\begin{center}
\includegraphics[width=\linewidth,clip]{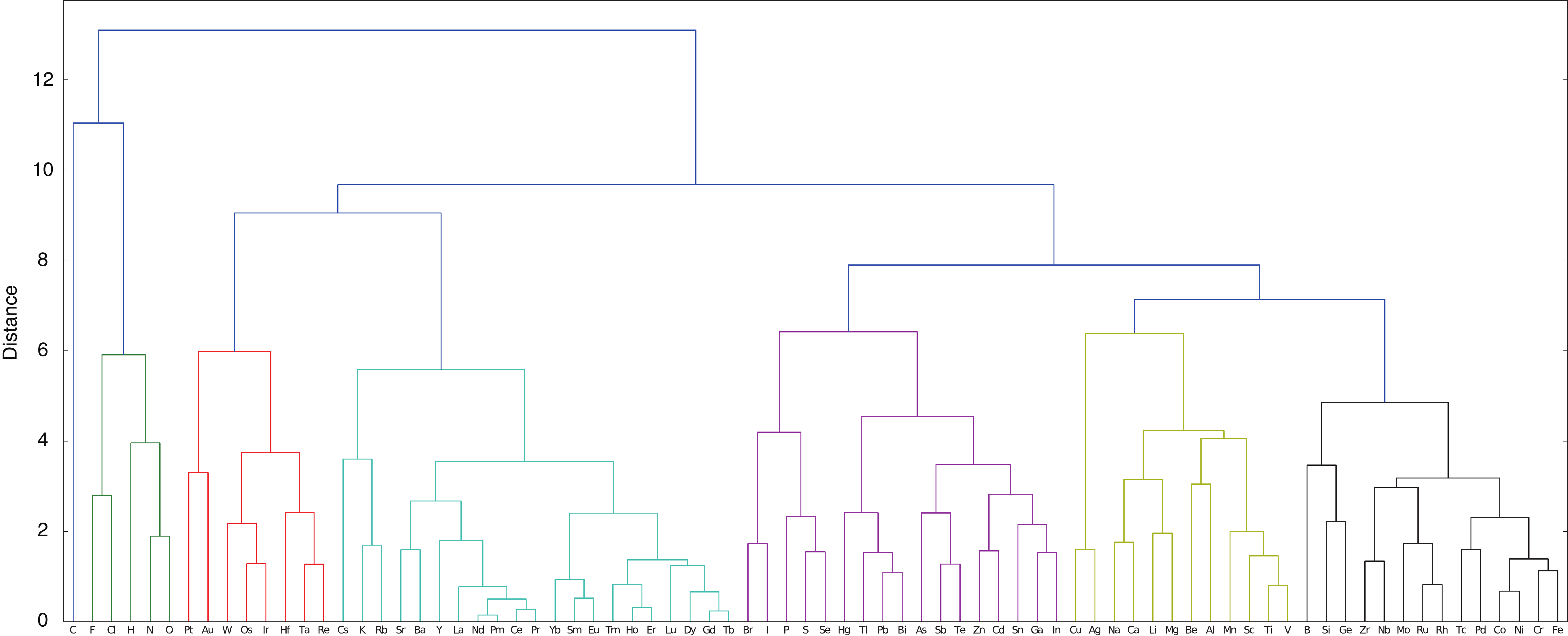} 
\caption{
Dendrogram for elements obtained from the complete-link clustering.
}
\label{recommend1:Fig7-Dendrogram}
\end{center}
\end{figure*}

In this study, the composition similarity is measured by the 165 descriptors generated from only 22 elemental representations.
Therefore, the composition similarity is related to the elemental similarity defined by the 22 elemental representations.
Figure \ref{recommend1:Fig6-PCAElement} shows the distribution of elements in the space of the first and second principal components (PCs), expressed by linear combinations of the 22 elemental representations.
Since the first and second PCs explain 65\% of the total variance of the distribution of elemental representations, the distance between two elements on this figure almost indicates the similarity between the two elements.
Figure \ref{recommend1:Fig7-Dendrogram} shows the dendrogram for elements obtained from a hierarchical clustering technique using the 22 elemental representations.
We perform the complete-link clustering, where the distance between clusters is the maximum distance between their members.

\section{Conclusion}
In this study, we report a machine-learning approach to discover currently unknown CRCs only from existing inorganic crystal structure databases.
Our approach is regarded as a descriptor-based recommender system for the discovery of new inorganic compounds.
The discovery rates of recommender systems are much higher than that of random sampling, demonstrating that the use of descriptors as a prior knowledge for compositional similarity significantly accelerates the discovery of currently unknown CRCs that are not present in the training database.
The most meaningful application of such descriptor-based recommender systems for currently unknown CRCs can be demonstrated in multicomponent systems, where only a small number of CRCs are known and a huge number of compositions can be candidates of CRCs. 
Therefore, the knowledge-based method of the present study should contribute significantly to the prediction of multicomponent currently unknown CRCs.

\begin{acknowledgments}
This work was supported by PRESTO, JST, and a Grant-in-Aid for Scientific Research on Innovative Areas ``Nano Informatics'' (Grant No. 25106005) from the Japan Society for the Promotion of Science (JSPS). 
AS, HH, and IT were also supported by ``Materials research by Information Integration'' Initiative (MI2I) from JST. 
IT was supported by a Grant-in-Aid for Scientific Research (A), JSPS (Grant No. 15H02286).
\end{acknowledgments}

\bibliography{recommend1}

\end{document}